\documentclass[aps,prc,groupedaddress,twocolumn,showpacs,amsmath,amssymb]{revtex4}
\usepackage{graphics}
\usepackage{dcolumn}
\usepackage{longtable}

\begin{document}

\title{Does deuteron-induced fission on actinide nuclei prevails over the breakup at low incident energies?}

\author{M.~Avrigeanu} \email{mavrig@ifin.nipne.ro}
\author{V.~Avrigeanu}
\affiliation{ Horia Hulubei National Institute for Physics and Nuclear Engineering, P.O. Box MG-6, 077125 Bucharest-Magurele, Romania}
\author{A.J. Koning}
\affiliation{Nuclear Research and Consultancy Group, P.O. Box 25, NL-1755 ZG Petten, The Netherlands}

\date{\today}

\begin{abstract}
An analysis of the $^{231}$Pa$(d,3n)$$^{230}$U reaction excitation function at energies around the Coulomb barrier has taken into account the pre--equilibrium and compound--nucleus cross sections corrected for the deuteron--breakup decrease of the total reaction cross section, as well as the inelastic breakup enhancement. The analysis reveals the dominance of the deuteron breakup mechanism unlike a former assessment in this respect of the deuteron--induced fission process.
\end{abstract}


\maketitle

\section{Introduction}
\label{Sec1}
The analysis of the deuteron--induced reactions at low energies, in terms of the usual nuclear reaction models, is challenging due to deuteron breakup (BU) following its weak binding energy, $B_d$=2.224 MeV. The various reactions initiated by the breakup neutrons and protons render the study and prediction of the deuteron reaction cross sections more complex. This is why recent measurements of the $^{231}$Pa$(d,3n)$$^{230}$U and $^{231}$Pa$(p,2n)$$^{230}$U reactions cross sections, between 11.2 and 19.9 MeV \cite{Pad_exp}, and respectively 10.6 and 23.8 MeV \cite{Pap_exp}, are particularly useful for the analysis of breakup effects on the former excitation function. The outgoing energy of the breakup--protons along the $^{231}$Pa$(d,3n)$$^{230}$U data of Ref. \cite{Pad_exp} is covered by the $^{231}$Pa$(p,2n)$$^{230}$U excitation function that can be used for the calculation of the inelastic BU enhancement of the $(d,3n)$ reaction cross sections. On the other hand, the following highlighting of the related BU effects, discussed at large elsewhere \cite{avrig09,bem09,avrig10,simek11,avrig11}, may show that there could be different ways to describe the same data \cite{Pad_exp}. While the deuteron--induced fission has previously been considered to be the dominant decay channel, the same role is attributed in this work to the deuteron BU. Consequently, we point out the need of additional measurements in order to establish which description is better. Nevertheless, a consistent analysis of the deuteron interactions is of real interest for applied objectives as the fusion technology or nuclear medicine \cite{Pad_exp} as well as for basic issues related to, e.g., the surrogate nuclear reaction method (\cite{sc10} and Refs. therein). Thus one may note the rising use of the $(d,xf)$ reaction, where $x$ stands for a proton or deuteron, as a surrogate for the $(n,f)$ reaction, and of $(d,p\gamma)$ as a surrogate for neutron capture. However, only the internal surrogate ratio method \cite{jma09} was shown to be valid in the presence of the deuteron breakup without assuming specific breakup reaction mechanisms.

\section{Deuteron breakup and induced fission}
\subsection{Deuteron breakup cross sections}
\label{Sec2}
The physical picture of the deuteron breakup in the Coulomb and nuclear fields of the target nucleus considers two distinct processes, namely the elastic breakup (EB) in which the target nucleus remains in its ground state and none of the deuteron constituents interacts with it, and the inelastic breakup or breakup fusion (BF), where one of these deuteron constituents interacts with the target nucleus while the remaining one is detected (e.g., \cite{avrig10} and Refs. therein). 
Under the assumption that the inelastic--breakup cross section for neutron emission $\sigma_{BF}^n$ is the same as that for the proton emission $\sigma_{BF}^p$ (e.g., Ref. \cite{must87}), the total breakup cross sections $\sigma_{BU}$is given by the sum
\begin{equation}\label{eq:1}
 \sigma_{BU} = \sigma_{EB} + 2\sigma_{BF}^{n/p} \:\:\:\:, 
\end{equation} 
while the total neutron-- and proton--emission breakup cross sections, $\sigma_{BU}^n$ and respectively $\sigma_{BU}^p$, are given by 
\begin{equation}\label{eq:2}
 \sigma_{BU}^{n/p} = \sigma_{EB} + \sigma_{BF}^{n/p} \:\:\:\:. 
\end{equation} 

On the other hand, empirical parameterizations have been established \cite{avrig09} for the total nucleon--emission breakup fraction
\begin{equation}\label{eq:3}
 f^{(n/p)}_{BU}=\sigma^{n/p}_{BU}/\sigma_R  \:\:\:\:, 
 \end{equation} 
and the elastic--breakup fraction
\begin{equation}\label{eq:4}
 f_{EB}=\sigma_{EB}/\sigma_R   \:\:\:\:, 
 \end{equation} 
where $\sigma_R$ is the deuteron total reaction cross section. 
Since a dependence of these fractions on atomic $Z$ and mass $A$ numbers of the target nucleus, and deuteron incident energy $E$ was found on the basis of experimental systematics \cite{avrig09}, the nucleon inelastic--breakup fraction may have the following form:
\begin{equation}\label{eq:7}
 f_{BF}^{(n/p)}=f^{(n/p)}_{BU}-f_{EB} \:\:\:\:, 
\end{equation} 
that leads to the nucleon inelastic--breakup cross sections:
\begin{equation}\label{eq:8}
\sigma^{n/p}_{BF} =  f^{(n/p)}_{BF} \sigma_R \:\:\:\:.
 \end{equation} 

\begin{figure} [b]
\resizebox{0.9\columnwidth}{!}{\includegraphics{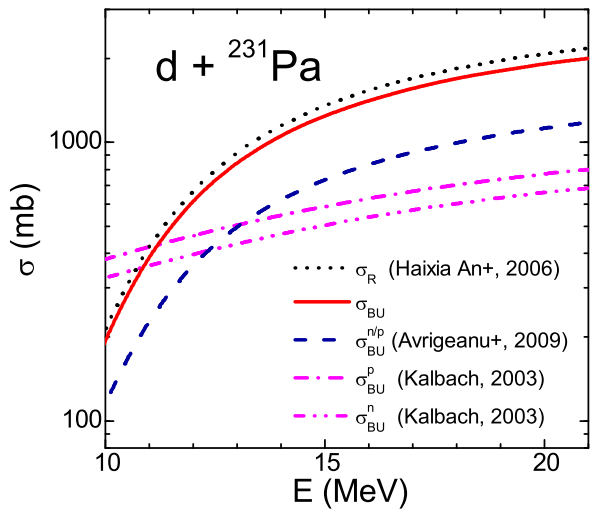}}
\caption{\label{Fig1}(Color online) Energy dependence of the total breakup (solid curve), total nucleon--emission breakup  cross sections (dashed) \cite{avrig09}, and total proton-- (dash-dotted curve) and neutron--emission breakup cross sections  (dash-dot-dotted) of Ref. \cite{kalb03}, for deuterons interacting with $^{231}$Pa around the Coulomb barrier. The deuteron total reaction cross sections \cite{ha06} are also shown (dotted curve).}
\end{figure}

However, for $A$$\sim$230 only the parametrization of the total nucleon--emission breakup fraction \cite{avrig09}
\begin{eqnarray}\label{eq:5}
f^{(n/p)}_{BU}=0.087-0.0066 Z + 0.00163 ZA^{1/3}+ \:  \nonumber \\
 0.0017A^{1/3}E-0.000002 ZE^2, 
\end{eqnarray}
%
could be accurate, due to the lack of related data for the elastic--breakup.  
In order to use the same approach for deuterons incident on the $^{231}$Pa target nucleus and the energy range of the data of Ref. \cite{Pad_exp}, we have taken into account the value $\sigma_{EB}$=119 mb that was considered for deuterons incident on the $^{232}$Th nucleus at $E$=15 MeV \cite{klein81} as well as its normalization corresponding to the measured proton inelastic--breakup cross section within the same work. Moreover, due to the lack of additional similar data, we used the resulted value $f_{EB}$=0.164 in the whole incident energy range, that is anyway rather narrow. Then we used the corresponding BU and BF components, given by Eqs. ($\ref{eq:5}$) and ($\ref{eq:7}$), to obtain the nucleon BF as well as the total BU cross sections shown in Fig.~\ref{Fig1}. In this respect a particular importance has the deuteron total reaction cross section in Eqs. ($\ref{eq:3}$) and ($\ref{eq:8}$), respectively, taken in the present work according to the RIPL-3 recommendation \cite{RIPL3} for the deuteron optical potential of Ref. \cite{ha06}. Actually this potential is the only one based on the data analysis for nuclei with $A$$\ge$208. The corresponding larger total reaction cross sections (Fig.~\ref{Fig1}) with respect to, e.g., the default deuteron potential within the computer code TALYS \cite{TALYS}, led to BU cross sections in the present work that are higher than the predictions of the Kalbach parameterization \cite{kalb03} for the total neutron-- and proton-emission breakup cross-sections that differ by normalization factors $K_{d,p/n}$:
\begin{equation}\label{eq:9}
\sigma^{n/p}_{BU}=K_{d,n/p}\frac{(A^{1/3}+0.8)^2}{1+exp\frac{(13-E)}{6}}, \:\:  K_{d,n} = 18, \:\: K_{d,p} = 21  \:\:,
\end{equation} 
that are also shown in Fig.~\ref{Fig1}. It results that, for deuteron incident energies above ${\sim}$13 MeV, the predictions for the total nucleon--emission breakup cross sections given by both parameterizations \cite{avrig09,kalb03} are around $\sim$50\% of the deuteron total reaction cross section. It is only the extrapolation of the Kalbach parameterization at low incident energies that leads to nucleon--emission BU cross sections exceeding even the deuteron total reaction cross section. Nevertheless, regardless of the differences between them, both parameterizations point out the dominance of the breakup mechanism at the deuteron incident energies below and around the Culomb barrier. Actually this conclusion is in line with the experimental total proton--emission BU fraction data for deuterons on $^{232}$Th \cite{wu79,klein81} that were taken into account within the above--mentioned systematics \cite{avrig09}.

\subsection{Fission competitive decay}
\label{Sec3}
\begin{figure} [t]
\resizebox{0.9\columnwidth}{!}{\includegraphics{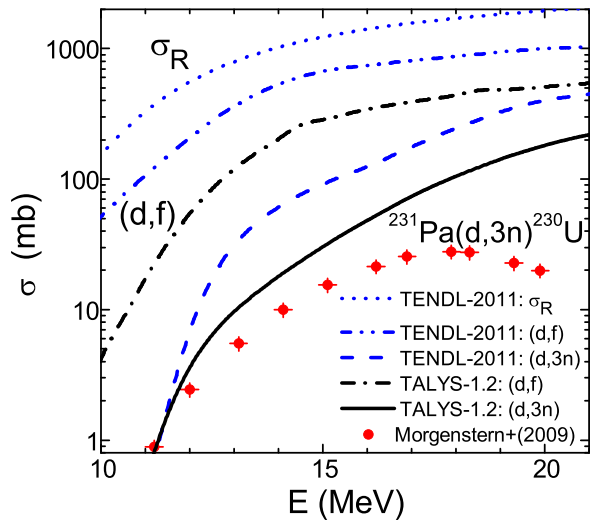}}
\caption{\label{Fig2}(Color online) Comparison of the excitation function of the reaction $^{231}$Pa(d,3n)$^{230}$U  measured \cite{Pad_exp}, calculated using the default options of the code TALYS \cite{TALYS} (solid curve), and most recently evaluated \cite{TENDL} (dashed). There are also shown the corresponding calculated (dash-dotted) and evaluated (dash-dot-dotted) cross sections for the deuteron--induced fission on $^{231}$Pa, together with the evaluated deuteron total reaction cross sections (dotted).}
\end{figure}

On the other hand, Morgenstern {\it et al.} \cite{Pad_exp} have found a definite dominance of the fission decay channel within their former analysis of the  $^{231}$Pa$(d,3n)$$^{230}$U reaction cross--section around the Coulomb barrier. They also noted, although without a quantitative assessment, that a significant decrease of the available compound--nucleus cross section occurs due to the deuteron breakup. Nevertheless, the fission cross section obtained within the EMPIRE-2 computer code assumptions \cite{EMPIRE} has been quite close to the deuteron total reaction cross section (Fig. 3 of Ref. \cite{Pad_exp}). Conversely, lower fission cross sections can be found straight away either using the code TALYS-1.2 with its default options \cite{TALYS} or within the TENDL-2011 library \cite{TENDL}, explicitly shown in Fig.~\ref{Fig2}. The two distinct sets of calculated results are used here just to prove the similar weight of the fission mechanism. The significant difference between them is due to the different optical potentials used for deuterons. 
Unfortunately, there are no $(d,f)$ measured cross sections but only an analysis of $^{231}$Pa$(d,pf)$ data at the deuteron energy of 15 MeV \cite{Back74}, where the measured fission probability versus the excitation energy of the residual nucleus $^{232}$Pa was found lower than 40$\%$ (Fig. 16 from Ref. \cite{Back74}). As a consequence, we focus on the measured $^{231}$Pa(d,3n)$^{230}$U excitation function in order to check the dominance of the breakup mechanism predicted by empirical parameterizations \cite{avrig09,kalb03}. 

\begin{figure} [b]
\resizebox{.95\columnwidth}{!}{\includegraphics{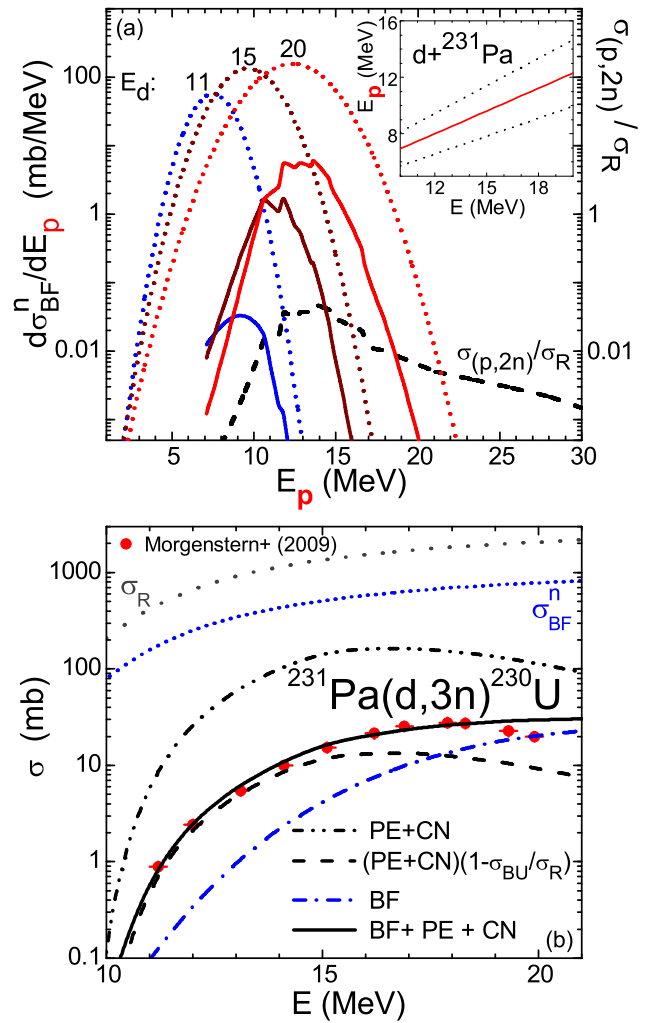}}
\caption{\label{Fig3}(Color online) (a) Results (solid curves) of the convolution of the cross section ratio $\sigma_{(p,2n)}$/$\sigma_{(p,R)}$ for the target nucleus $^{231}$Pa (dashed) and the Gaussian distribution (dotted) of breakup--protons energies for deuterons on $^{231}$Pa at incident energies of 10, 15 and 20 MeV noted on their top; in the insert: centroid of the Gaussian distribution of breakup--protons energies \cite{kalb08} versus the deuteron incident energy (solid curve) on $^{231}$Pa, and the related $E_p\pm\Gamma/2$ values (dashed). (b) As for Fig.~\ref{Fig2} but for deuteron total reaction cross sections \cite{ha06} (dotted), nucleon inelastic--breakup cross section (short dotted), inelastic BU enhancement (dash-dotted), and the $(d,3n)$ reaction cross sections calculated without (dash-dot-dotted) and with (dashed) BU decrease of $\sigma_R$, as well as including the BU enhancement (solid).} 
\end{figure}

\section{Detailed breakup mechanism}
\label{Sec4}
The dominance of the breakup mechanism has two opposite effects. Firstly, the deuteron total breakup cross section reduces significantly the amount of total reaction cross section that should be shared among different outgoing channels. This effect is shown in Fig.~\ref{Fig3}(b) for the  $^{231}$Pa$(d,3n)$$^{230}$U reaction and the calculated cross sections using the computer code TALYS-1.2 \cite{TALYS}. In order to emphasize the two distinct BU effects, we made a different choice than for the results shown in Fig.~\ref{Fig2}. Subsequently we have not used the BU--inclusion option of TALYS by means of the Kalbach parameterization \cite{kalb03}. Thus, we have obtained firstly the pre--equilibrium (PE) and compound nucleus (CN) contributions to the $(d,3n)$ reaction cross sections, under the assumption of no breakup process. Then the BU reduction of these results was addressed by using a reduction factor $(1-\sigma_{BU}/\sigma_R)$ of the deuteron total reaction cross section. The $(d,3n)$ reaction cross sections obtained in this way are now in good agreement with the measured data just above the effective reaction threshold while formerly these data 
were also greatly overestimated. However, an underestimation by a factor up to 3 at $E$$\sim$20 MeV becomes visible in Fig.~\ref{Fig3}(b). Nevertheless, the description of the reaction cross sections at the lowest energies may validate the PE and CN model parameters used in these calculations. Apart from the default TALYS parameters, we used the above--mentioned deuteron optical potential \cite{ha06}, the nucleon optical potentials for actinides \cite{rc08} from the RIPL 2408 and RIPL 5408 potential segments \cite{RIPL3} for neutrons and protons, respectively, the microscopic level densities of Hilaire {\it et al.} \cite{sg08}, and the WKB approximation for the fission path model \cite{ms06}.

Secondly, we aim to account for the inelastic breakup enhancement due to one of the deuteron constituents that interacts with the target nucleus and leads to a secondary composite nucleus, with further significant contributions to various primary reaction channels. The secondary-- and third--chance particle emissions next to the original deuteron--target interaction \cite{bem09,simek11,avrig11} are thus especially enhanced. In the present case, the absorbed proton following the breakup neutron emission contributes to the enhancement of the $^{230}$U activation cross section through the $^{231}$Pa$(p,2n)$$^{230}$U reaction. 

In order to calculate the breakup enhancement for the  $^{231}$Pa$(d,3n)$$^{230}$U reaction, the nucleon inelastic--breakup cross section $\sigma_{BF}^n$ given by Eq. ($\ref{eq:8}$) was formerly considered \cite{avrig11} together with the ratio $\sigma_{(p,2n)}$/$\sigma_{(p,R)}$ that corresponds to the weight of the above--mentioned reaction induced by the breakup--protons on the $^{231}$Pa target nucleus \cite{avrig11}. We used in this respect the measured $^{231}$Pa$(p,2n)$$^{230}$U reaction cross sections \cite{Pap_exp} and the proton total reaction cross section $\sigma_{(p,R)}$ generated by the above--mentioned optical potential. We may express this ratio as a function of the deuteron incident energy, using the recent Kalbach \cite{kalb08} formula for the center--of--mass system centroid of the Gaussian distribution of breakup--proton energies. Since the breakup--proton energy range of $\sim$9--14 MeV corresponds to the incident energies of 11--20 MeV of the measured $(d,3n)$ excitation function \cite{Pap_exp}, the $\sigma_{(p,2n)}$ cross sections values are provided just by the measurements, no other reaction model calculations being involved. This is why the simultaneous analysis of $(d,3n)$ \cite{Pad_exp} and $(p,2n)$ experimental excitation functions \cite{Pap_exp} is so useful for the study of the inelastic breakup and complementary reaction mechanisms considered for the deuteron interactions with nuclei. 

However, a better estimation for the inelastic breakup enhancement is given by the convolution of the ratio $\sigma_{(p,2n)}$/$\sigma_{(p,R)}$ and the Gaussian distribution of the breakup--proton energies corresponding to a given incident deuteron energy  \cite{kalb08}. The former as well as the latter quantities for three deuteron incident energies are shown in Fig.~\ref{Fig3}(a). The same method has previously been applied \cite{simek11,avrig11} using a former Gaussian line shape \cite{kalb03}. The areas of the related convolution results  correspond to the inelastic--breakup enhancement of the $(d,3n)$ reaction cross sections at the given deuteron energies. The energy dependence of this inelastic--breakup enhancement of the $^{231}$Pa$(d,3n)$$^{230}$U activation cross section is shown in Fig.~\ref{Fig3}(b), while the corresponding total activation of $^{230}$U is finally compared with the experimental data \cite{Pad_exp}. As expected, the more realistic treatment of the inelastic breakup enhancement by taking into account the quite large widths $\Gamma$ of the breakup--proton energy distributions (see the upper insertion in Fig. 3(a)), has led to a rather accurate description of data . Further improvements of the breakup analysis could add to a better account of the related energy dependence.\\

\section{Conclusions}
\label{Sec5}
An analysis of the $^{231}$Pa$(d,3n)$$^{230}$U reaction excitation function at energies around the Coulomb barrier has taken into account the pre--equilibrium and compound--nucleus cross sections corrected for the deuteron--breakup decrease of the total reaction cross section, as well as the inelastic breakup enhancement. The analysis reveals the dominance of the deuteron breakup mechanism unlike a former assessment in this respect of the deuteron--induced fission process.

The improvement of deuteron breakup effects estimation requires  complementary experimental studies of, e.g., the $(d,3n)$ and $(p,2n)$ reaction cross sections for the same target nucleus and within correlated incident--energy ranges. The suitability of the empirical parameterization of the breakup components and reaction mechanisms involved in the interaction process could thus be checked and updated. Furthermore, the associated inclusive neutrons and protons spectra measurements that allow the distinction among various contributing mechanisms are highly requested too, as well as $(d,pf)$ angular correlations when the deuteron induced fission process is analyzed. Given the increased interest in surrogate reaction studies, e.g. Ref. \cite{bj11}, the usefulness of detailed theoretical and experimental investigations of the breakup of weakly bound projectiles including deuterons is obvious.

\section*{Acknowledgments}
This work was partly supported by the grant of the Romanian National Authority for Scientific Research, CNCS-–UEFISCDI, project No. PN-II-ID-PCE-2011-3-0450.

\end{document}